\documentclass{dcds-B} 
 
\usepackage{epsfig} 
 
\def\currentvolume{??} 
\def\currentissue{??} 
\def\currentyear{??} 
\def\currentmonth{??} 
\def\ppages{1--17} 
 
\newcommand{\fr}[2]{\frac{\displaystyle{#1}}{\displaystyle{#2}}}

\title[Parametric dependences of non-linear singular maps] 
{\bf On the parametric dependences of a class of non-linear singular maps} 
\author[T. Gilbert and J. R. Dorfman]{} 
 
 
\email{thomas.gilbert@inln.cnrs.fr} 
\email{jrd@ipst.umd.edu} 
 
\begin{document} 
 
\maketitle 
 
\centerline{\scshape T. Gilbert\dag\ddag and J. R. Dorfman\dag} 
\medskip 
{\footnotesize \centerline{\dag\ Department of Physics and}  
\centerline{Institute for 
    Physical Science and Technology}  
\centerline{University of Maryland} 
\centerline{College Park, MD 20742, USA}} 
\medskip 
{\footnotesize \centerline{\ddag\ Laboratoire Cassini, CNRS,} 
\centerline{Observatoire de la C\^ote d'Azur} 
\centerline{B. P. 4229, 06304 Nice, France} 
\centerline{and}
\centerline{Institut Non-Lin\'eaire de Nice}
\centerline{CNRS, Universit\'e de Nice}  
\centerline{1361 Route des Lucioles, 06560 Valbonne, France}}
\medskip 
 
 
\medskip 
 
\begin{abstract} 
We discuss a two-parameter family of maps that generalize piecewise linear, 
expanding maps of the circle. One parameter measures the effect of a 
non-linearity which bends the branches of the linear map. The second 
parameter rotates points by a fixed angle. For small values of  
the nonlinearity parameter, we compute the invariant measure and show that 
it has a singular density to first order in the nonlinearity parameter. 
Its Fourier modes have forms similar to the Weierstrass 
function. We discuss the consequences of this singularity on the Lyapunov  
exponents and on the transport properties of the corresponding multibaker  
map. For larger non-linearities, the map becomes non-hyperbolic  
and exhibits a series of period-adding bifurcations. 
\end{abstract} 
 
\section{Introduction} 
 
The concept of natural invariant measure is central to the theory of 
Sinai-Ruelle-Bowen measures. Rigorous results pertaining to the existence 
and uniqueness of such measures 
have been obtained for the most part in the framework of Anosov 
diffeomorphisms and Axiom A systems. However, as discussed by Chernov 
\cite{chernov02}, there are hyperbolic maps other than diffeomorphisms for 
which SRB measures were constructed \cite{Pes92,Y98,CE01}. It is our purpose to 
investigate some of the properties of a class of such systems with 
singularities. 
 
This paper is concerned with the parametric dependence of a particularly 
simple class of one-dimensional circle maps which may be regarded as  
non-linear generalizations of a class of piece-wise linear, expanding maps. The 
generalized maps have two essential properties, 
the first being that these maps are singular in the sense that their first 
derivative posseses a singularity (or a finite set of them), and second 
that the maps perform some irrational rotation such that the singularity is 
moved ergodically about the circle upon iteration. 
This class of maps will be characterized by two parameters: a 
first parameter which measures the strength of the nonlinearity, and a 
second one which is a 
rotation angle.  
 
Depending on the strength of the non-linearity, the map 
can be {\em hyperbolic} or be {\em non-hyperbolic}. In the framework of 
the one-dimensional circle maps we consider here, these two regimes 
are best described by measures on interval-filling attractors or by 
delta-measures, respectively. It is convenient  
to assume that the linear regime, i.~e. when the non-linearity is set to 
zero, corresponds to a regime where the map is uniformly 
expanding, for instance, an 
angle-doubling automorphism. The onset of non-linearity perturbs this 
hyperbolic regime without altering this property. However, when the 
non-linearity reaches a threshold value, the map loses its hyperbolicity 
and stable periodic orbits may occur. 
 
The presence of a rotation has interesting consequences in both these 
regimes. In the regime of small non-linearity, the hyperbolicity will not 
be affected by the rotation since the map is everywhere expanding. However, 
because of the existence of singularities, upon iteration, an initially 
uniform density will evolve into a discontinuous one.  
The invariant density has a dense set of discontinuities, but remains 
finite everywhere. For larger non-linearity, the rotation can take a 
non-expanding region into an expanding one, with the consequence that the 
attractor may alternate between hyperbolic and non-hyperbolic regions as 
the intensity of the non-linearity or the rotation angle are varied. 
 
\section{\label{SMsrb}Singular circle maps} 
 
Consider the family of automorphisms of the circle  
\begin{equation} 
\varphi_{0,\zeta}\::\:x\longrightarrow 2 x + \zeta \mod 1\ , 
\label{eq0} 
\end{equation} 
which combines a rotation by an angle $\zeta$ with a uniformly expanding 
automorphism. It is clear that the Lebesgue measure is invariant under such 
a map, regardless of the parameter $\zeta$. The parametric dependences of 
maps with a rotation become less trivial when the linear expansion is 
replaced by a nonlinear one. Consider maps of the form 
\begin{equation} 
\varphi_{a,\zeta}\::\:x\longrightarrow 2 f_a(x) + \zeta \mod 1\ , 
\label{eq1} 
\end{equation} 
where we assume $f_a$ is defined over the unit 
interval, with the following features, motivated by previous studies 
of non-linear baker maps\cite{gilbert99} ~: 
\begin{itemize} 
\item[(i)] 
the end points $0,1$ of the interval are fixed points, $f_a(0) = 0$ and $f_a(1) = 1$,  
such that the former is an attracting fixed point, $f_a'(0)<1$, while the latter is 
repelling, $f_a'(1)>1$; 
\item[(ii)] $f_a$ is analytic with respect to the parameter $a$; 
\item[(iii)] $f_a(x)$ is Holder continous with respect to the variable 
  $x$. Moreover, all the derivatives of $f_a$ with respect to $x$ exist at the 
  end points; 
\item[(iv)] $f_a$ has a semi-group property, namely the composition of $f_a$ 
  with itself yields $f_{2 a}$,  
\begin{equation} 
f_a\circ f_a(x) = f_{2 a}(x)\label{addfa}\ , 
\end{equation} 
and the inverse of $f_a$ is obtained by changing the sign of $a$, 
\begin{equation} 
f_a^{-1}(x) = f_{-a}(x)\ ; 
\label{invfa} 
\end{equation} 
\item[(v)]  
$f_a$ has the following property of {\em 
  reversibility} 
\begin{equation} 
f_a(1-x) = 1 - f_{-a}(x)\ . 
\label{trfa} 
\end{equation} 
\end{itemize} 
 
The semi-group property, Eqs. (\ref{addfa}-\ref{invfa}), imply that $f_a$ 
is the identity for $a=0$, namely $f_0(x) = x$; 
 
We note that Eqs. (\ref{invfa}-\ref{trfa}),  
imply the following~: 
\begin{equation} 
f_a'(0) f_a'(1)=1\ . 
\label{singularity} 
\end{equation} 
Moreover these quantities are different from $1$ as  
required in item (i). 
 
In this paper, we will consider in its generality the class of  
perturbations $f_a$ satisfying the above properties. In particular, we will 
consider the small parameter regime, $|a|\ll 1$, where $f_a$ can be written 
as a power expansion in its parameter, 
\begin{equation} 
f_a(x) = x + a f^{(1)}(x) + O(a^2), 
\label{faexp} 
\end{equation} 
and demand that the property of the derivative of $f_a$ at the extremities of 
the interval, see item (i) above,  hold to first order in $a$. 
 
An immediate property of $f^{(1)}$ follows from Eq. (\ref{trfa}), namely  
it is symmetric with respect to $x=1/2$,  
\begin{equation} 
f^{(1)}(x)= f^{(1)}(1 - x), 
\label{trf1} 
\end{equation} 
which also implies that the derivatives of $f^{(1)}$ at $x=0$ and $x=1$ 
have opposite slopes, 
\begin{equation} 
{f^{(1)}}'(0) = - {f^{(1)}}'(1). 
\label{f1der} 
\end{equation} 
Therefore the perturbations $f_a$ under consideration have a singularity of  
order $a$ in their derivatives at the end points of the unit interval, which 
are identified by the modulo one operation. 
 
Possible examples of such perturbations are polynomials or trigo\-nometric   
functions. Such examples are 
\begin{eqnarray} 
f_a(x) &=& x + a x(1 - x) + O(a^2)\ ,\label{ex1}\\ 
f_a(x) &=& x - \frac{a}{\pi}\sin(\pi x) + O(a^2)\ .\label{ex2} 
\end{eqnarray} 
\begin{figure}[phtb] 
\centerline{\psfig{figure=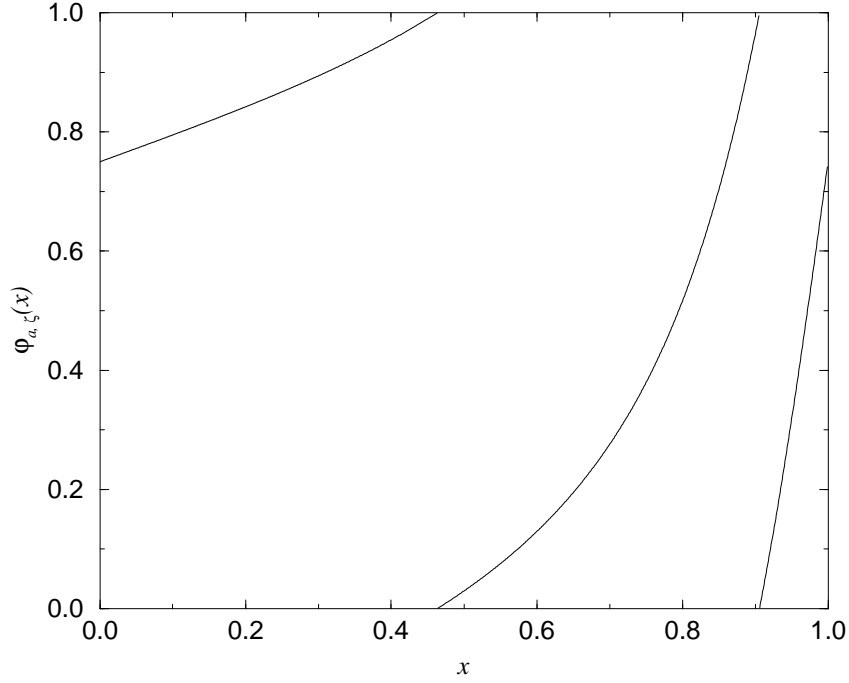,width=13cm}} 
\caption{\label{figSM2}{Illustration of the map 
    $\varphi_{a,\zeta}$ defined by Eq. (\ref{eq1})}, with $f_a(x)$  
  the full form of Eq. (\ref{ex2}) given later in 
  Eq. (\ref{SMcurtaina}). Here $a = 1.5$ and $\zeta = 0.75$.} 
\end{figure} 
\section{Hyperbolic regime} 
 
Although $\varphi_{a,\zeta}$ is a singular map, it has a unique absolutely  
continuous invariant measure, provided it is everywhere expanding, that is 
\begin{equation} 
2 f_a'(x) > 1. 
\label{expanding} 
\end{equation} 
The density of this  
measure has very interesting properties. Indeed, 
a direct consequence of the singularity of $f_a$ at the origin is that the 
iteration of a density of points initially uniform on the interval will yield  
a discontinuity which, provided $a$ is small, is of the order of $a$.  
At the next iteration this discontinuity is rotated around  
the circle by $\zeta.$ Further iterations yield more and more  
discontinuities which, as long as $\zeta$ is irrational, will cover densely  
the whole interval.  
 
The invariant density is a fixed point of the Frobenius-Perron operator, 
\begin{equation} 
\rho(x) =  
\left\{ 
\begin{array}{l@{\quad}l} 
\frac{1}{2}f_{-a}'\left(\frac{x - \zeta}{2} + 1\right) 
\rho\left[f_{-a}\left(\frac{x - \zeta}{2} + 1\right)\right]\\ 
\\ 
\quad + \frac{1}{2}f_{-a}'\left(\frac{x + 1 - \zeta}{2}\right) 
\rho\left[f_{-a}\left(\frac{x + 1 - \zeta}{2}\right)\right], 
&0\leq x <\zeta,\\ 
\\ 
\frac{1}{2}f_{-a}'\left(\frac{x - \zeta}{2}\right) 
\rho\left[f_{-a}\left(\frac{x - \zeta}{2}\right)\right]\\ 
\\ 
\quad + \frac{1}{2}f_{-a}'\left(\frac{x + 1 - \zeta}{2}\right) 
\rho\left[f_{-a}\left(\frac{x + 1 - \zeta}{2}\right)\right], 
&\zeta\leq x <1, 
\end{array} 
\right. 
\label{rho} 
\end{equation} 
where we dropped the parametric dependences to simplify the notation. 
 
Let us consider the first order term in the expansion of $\rho$ in $a$,  
namely 
\begin{equation}\label{SMrhox} 
\rho(x) = 1 + a\rho^{(1)}(x) + O(a^2)\ , 
\end{equation} 
where we have taken the uniform density, $\rho(x) = 1$ to be the 
absolutely continuous 
solution of Eq. (\ref{rho}) when $a=0$. 
An expression for $\rho^{(1)}$ follows from Eq. (\ref{rho})~:  
\begin{equation}\label{SMPFxmap} 
\rho^{(1)}(x) = \left\{ 
\begin{array}{ll} 
-\fr{1}{2}{f^{(1)}}'\left(\fr{x - \zeta }{2} + 1\right) -  
\fr{1}{2}{f^{(1)}}'\left(\fr{x + 1 - \zeta}{2}\right) &\\  
\\ 
\quad+\fr{1}{2}\rho^{(1)}\left(\fr{x - \zeta}{2} + 1\right) + 
\fr{1}{2}\rho^{(1)}\left(\fr{x + 1 - \zeta}{2}\right),& 
0 \leq x < \zeta,\\ 
\\ 
-\fr{1}{2}{f^{(1)}}'\left(\fr{x - \zeta }{2}\right) -  
\fr{1}{2}{f^{(1)}}'\left(\fr{x + 1 - \zeta}{2}\right) &\\ 
\\ 
\quad+\fr{1}{2}\rho^{(1)}\left(\fr{x - \zeta}{2}\right) + 
\fr{1}{2}\rho^{(1)}\left(\fr{x + 1 - \zeta}{2}\right),& 
\zeta \leq x < 1. 
\end{array} 
\right. 
\end{equation} 
 
Let us comment here that $\rho^{(1)}$ is generally not expected to be a 
well-behaved function of its arguments. This is readily seen if one notices 
that the two RHS of Eq. (\ref{SMPFxmap}) evaluate to different values at $x 
= \zeta$.  This is due to the property Eq. (\ref{f1der}) that the 
derivative of $f^{(1)}$ is discontinuous at the origin.  
Moreover the discontinuity is of order of 
$\delta\equiv\mathcal{O}[{f^{(1)}}'(0)]$. Similarly, at $x=3\zeta\mod 1$ --~the 
image of $x=\zeta$ under $\varphi_{0,\zeta}$, Eq. (\ref{eq0})~-- another 
discontinuity appears,  
this time of order $\delta/2$. In general, a discontinuity of order 
$\delta/2^n$ occurs at the $n$th iterate of $\zeta$ under 
$\varphi_{0,\zeta}$, i.~e. at $x = (\sum_{i=0}^n 2^i)\zeta\mod 1$. 
For rational values of $\zeta$, its binary expansion is a finite series~: 
there exists a finite integer $N$ such that $\zeta = \sum_{i=0}^N 
\omega_i/2^{i+1}$, $\omega_i\in\{0,1\}$ and $\omega_N=1$. In that  
case, after $N-1$ iterations of $\varphi_{0,\zeta}$, $\zeta$ is mapped back 
on one of the iterates already visited, namely the one corresponding to the 
next largest index $i$ for which $\omega_i=1$. Therefore, for rational 
$\zeta$'s, the set of discontinuities is finite. However this is not so for 
irrational $\zeta$'s. In that case, the set of discontinuities is dense, even 
though of exponentially small amplitudes.  
 
Hence $\rho^{(1)}$ is a distribution, the formal expression of which can be  
be found by representing it as a Fourier series~:  
\begin{equation}\label{SMFourier} 
\rho^{(1)}(x) = \sum_{k = 1}^\infty \left[ A_k(\zeta)\cos(2\pi k x) +  
B_k(\zeta)\sin(2\pi k x)\right]. 
\end{equation} 
The coefficients $A_k(\zeta)$ and $B_k(\zeta)$ satisfy the recursion relations  
\begin{eqnarray} 
A_k(\zeta) &=& -g_k \sin(2\pi k\zeta) 
+ A_{2k}(\zeta)\cos(2\pi k\zeta) - B_{2k}(\zeta)\sin(2\pi k\zeta), 
\nonumber\\ 
\label{SMrecrel1}\\ 
B_k(\zeta) &=& g_k \cos(2\pi k\zeta) 
+ A_{2k}(\zeta)\sin(2\pi k\zeta) + B_{2k}(\zeta)\sin(2\pi k\zeta), 
\nonumber\\ 
\label{SMrecrel2} 
\end{eqnarray} 
with solutions 
\begin{eqnarray} 
A_k(\zeta) &=& 
- \sum_{n = 0}^\infty g_{2^n k}\sin\left[2(2^{n + 1} - 1)\pi k \zeta\right], 
\label{SMAk}\\ 
B_k(\zeta) &=&  
\sum_{n = 0}^\infty g_{2^n k}\cos\left[2(2^{n + 1} - 1)\pi k \zeta\right], 
\label{SMBk} 
\end{eqnarray} 
where  
\begin{equation} 
g_k = 8\pi k\int_0^1 dx f^{(1)}\left(\frac{x}{2}\right)\cos(2\pi k x ). 
\label{gk} 
\end{equation}  
It is not difficult to show that $g_k$ decays like $1/k$ for $k\gg 1$. Indeed provided all the 
derivatives of $f^{(1)}$ exist at $x=0$, the integral on 
the RHS of Eq. (\ref{gk}) can be written in the form of a series in $1/k^{2 
  n}$ ($n$ integer) with bounded coefficients.  
 
\begin{figure}[phtb] 
\centerline{\psfig{figure=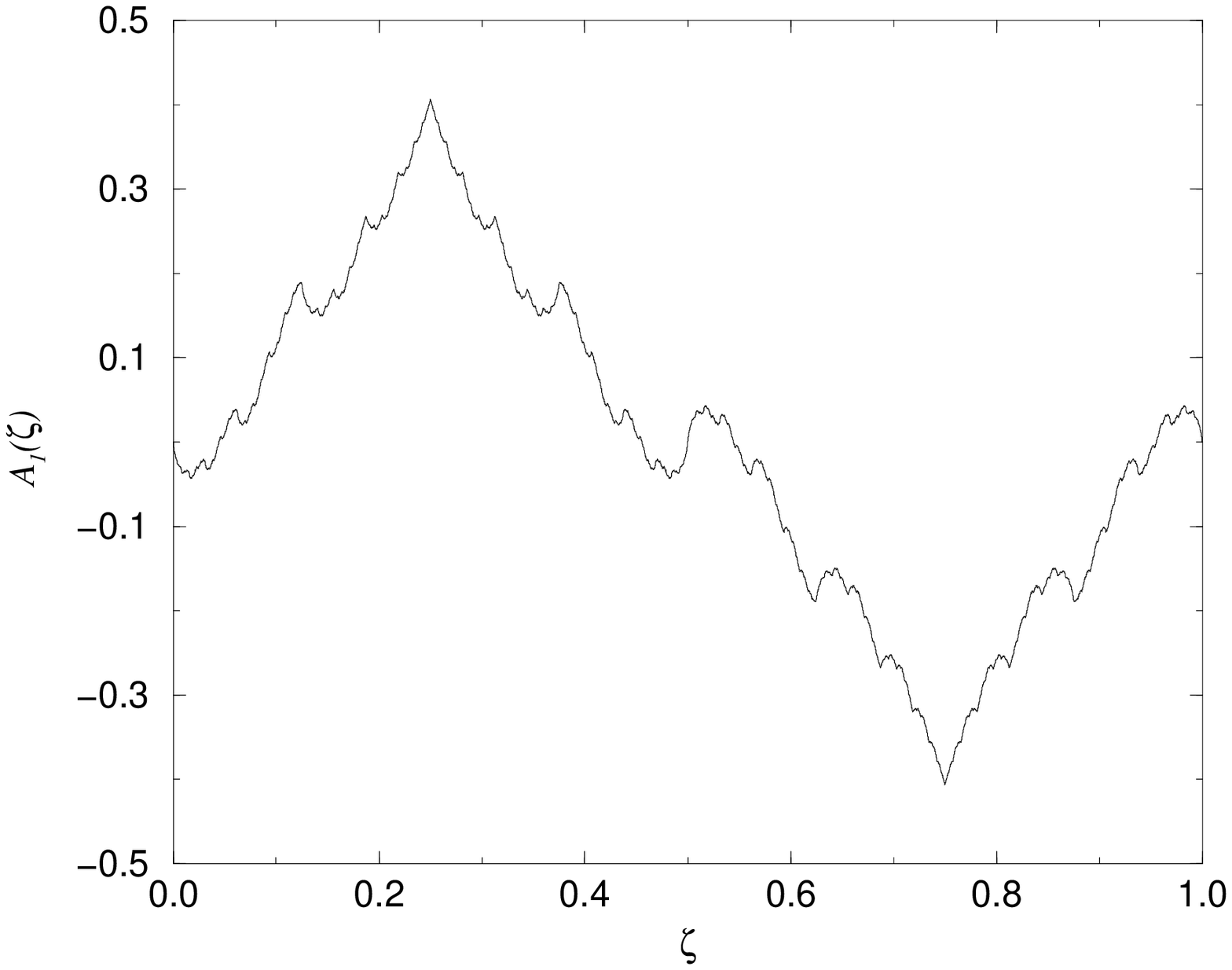,width=7cm} 
\psfig{figure=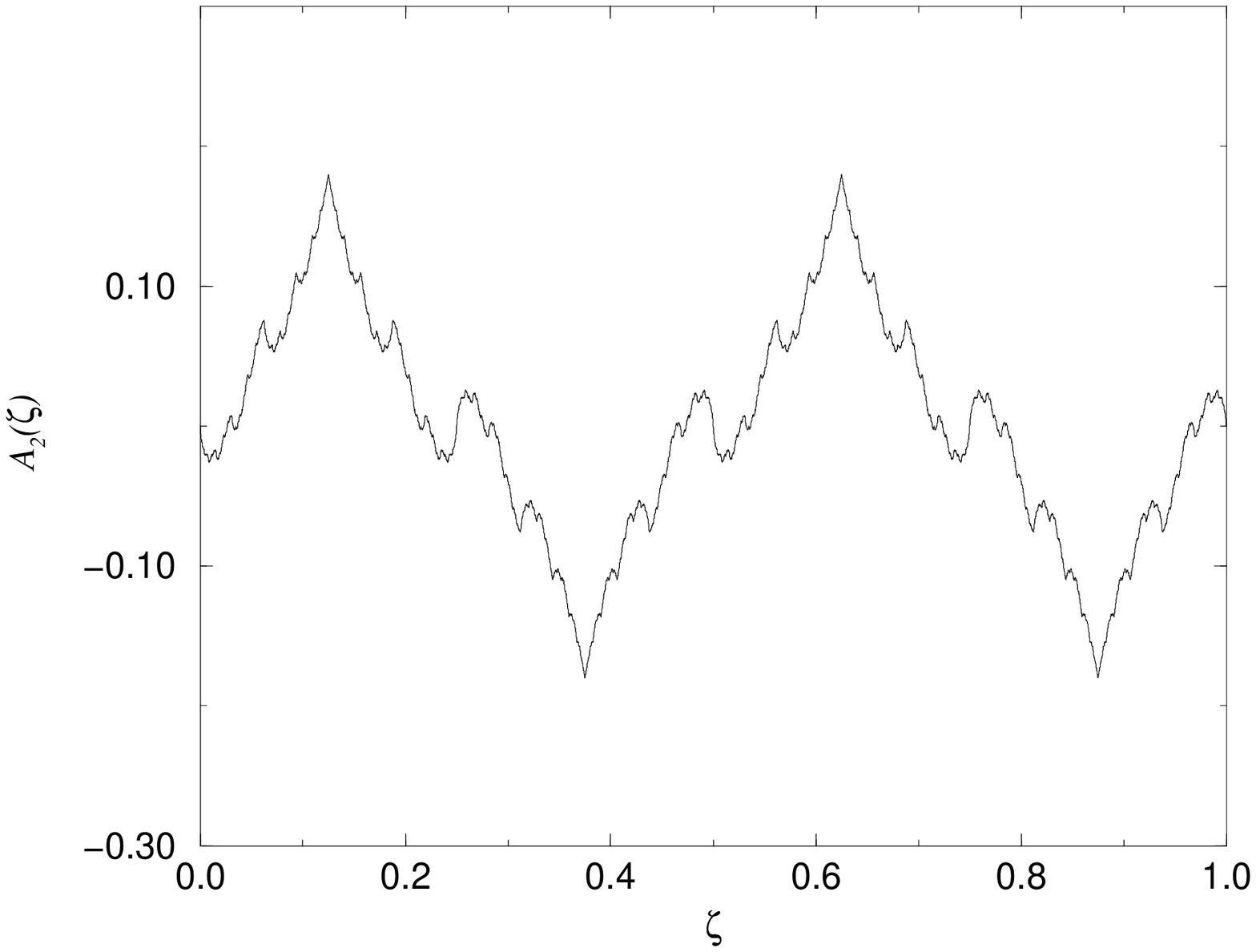,width=7cm}} 
\centerline{\psfig{figure=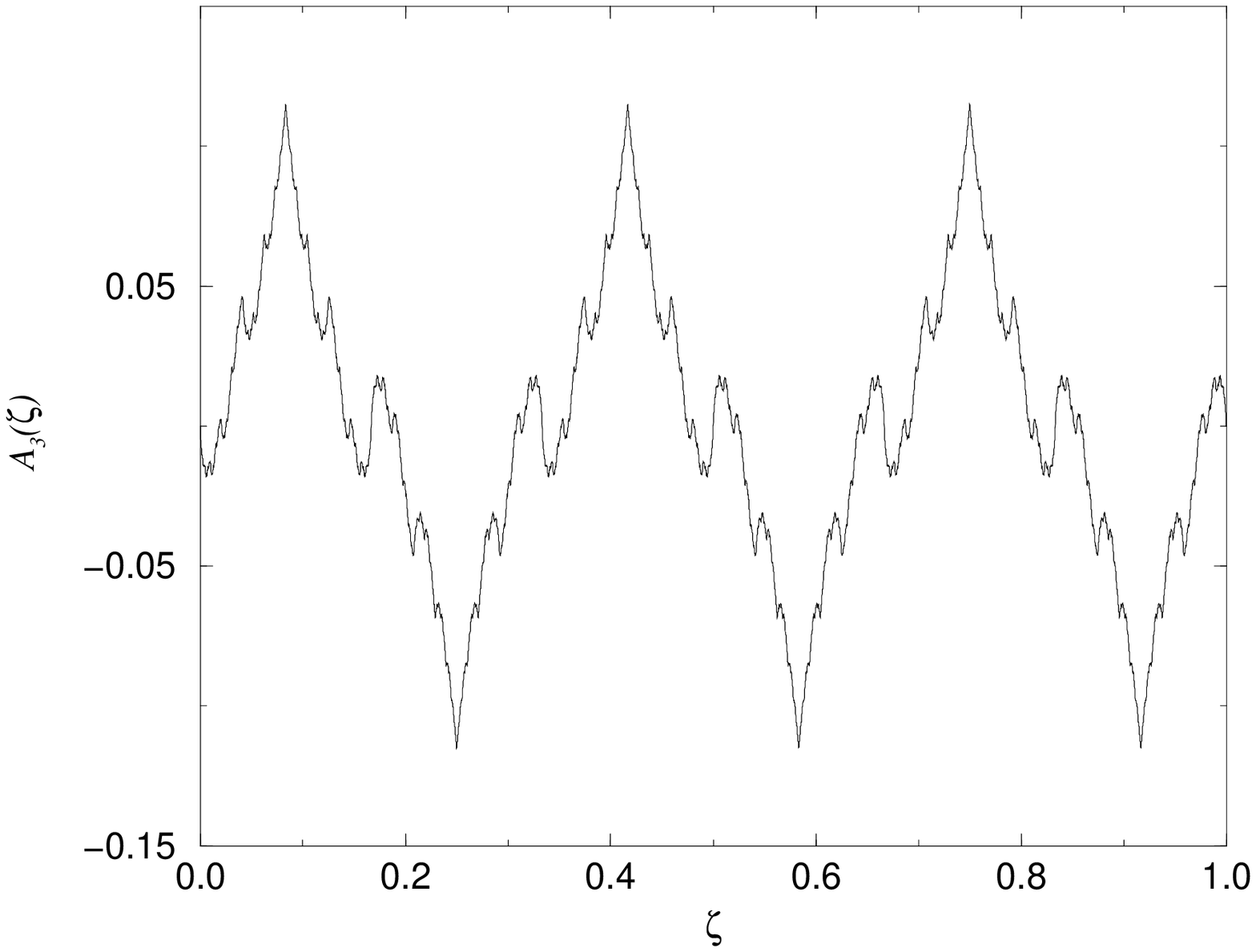,width=7cm} 
\psfig{figure=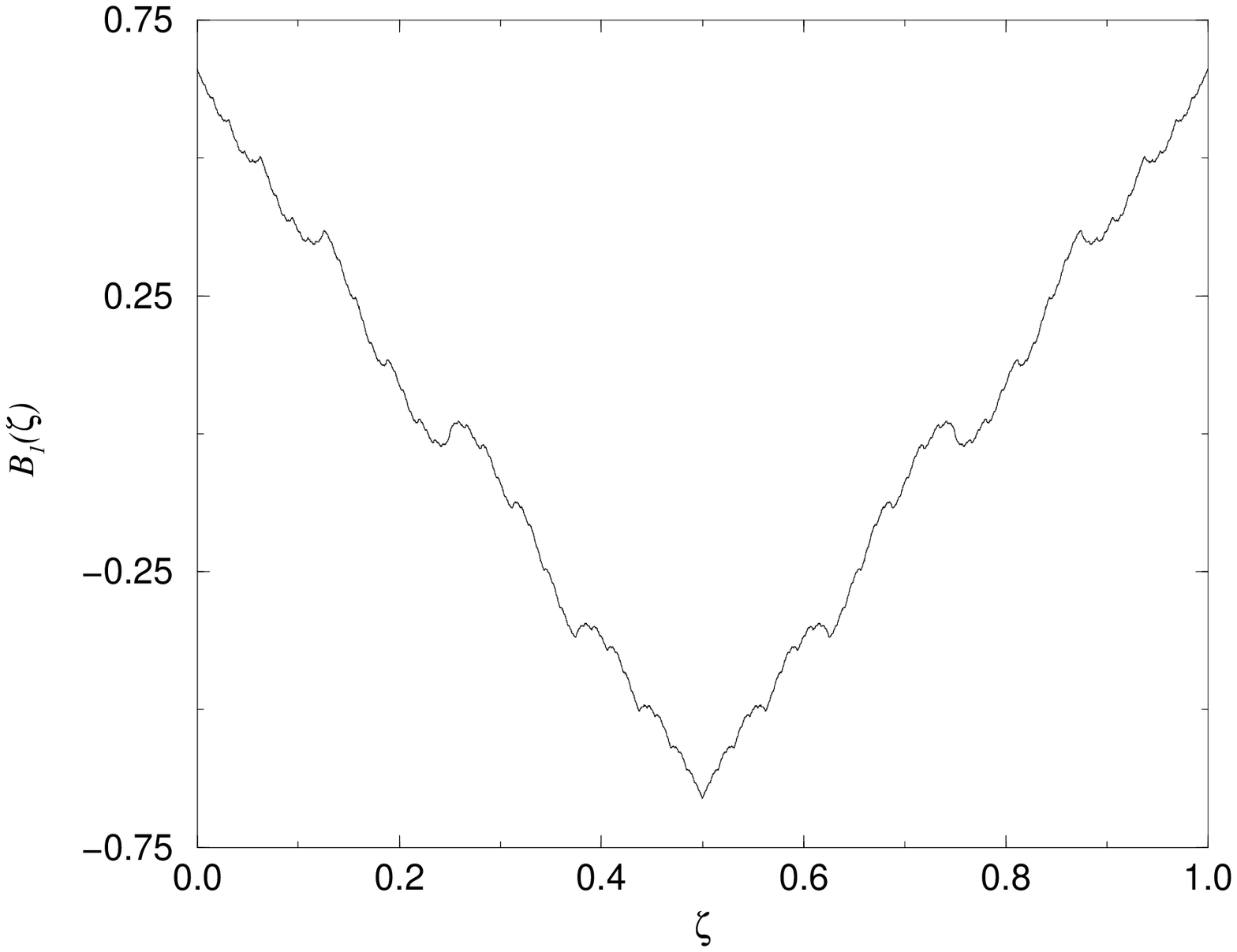,width=7cm}} 
\centerline{\psfig{figure=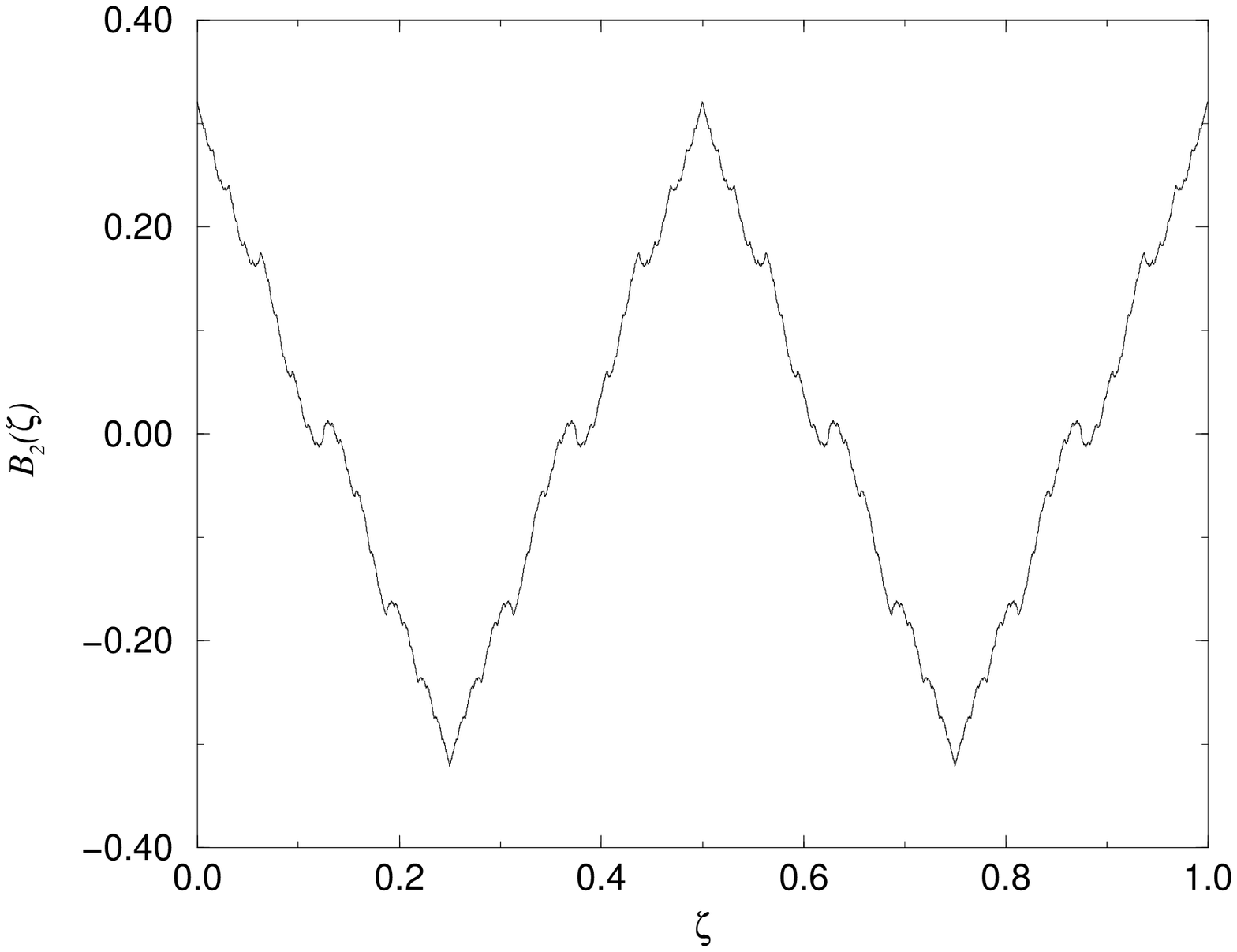,width=7cm} 
\psfig{figure=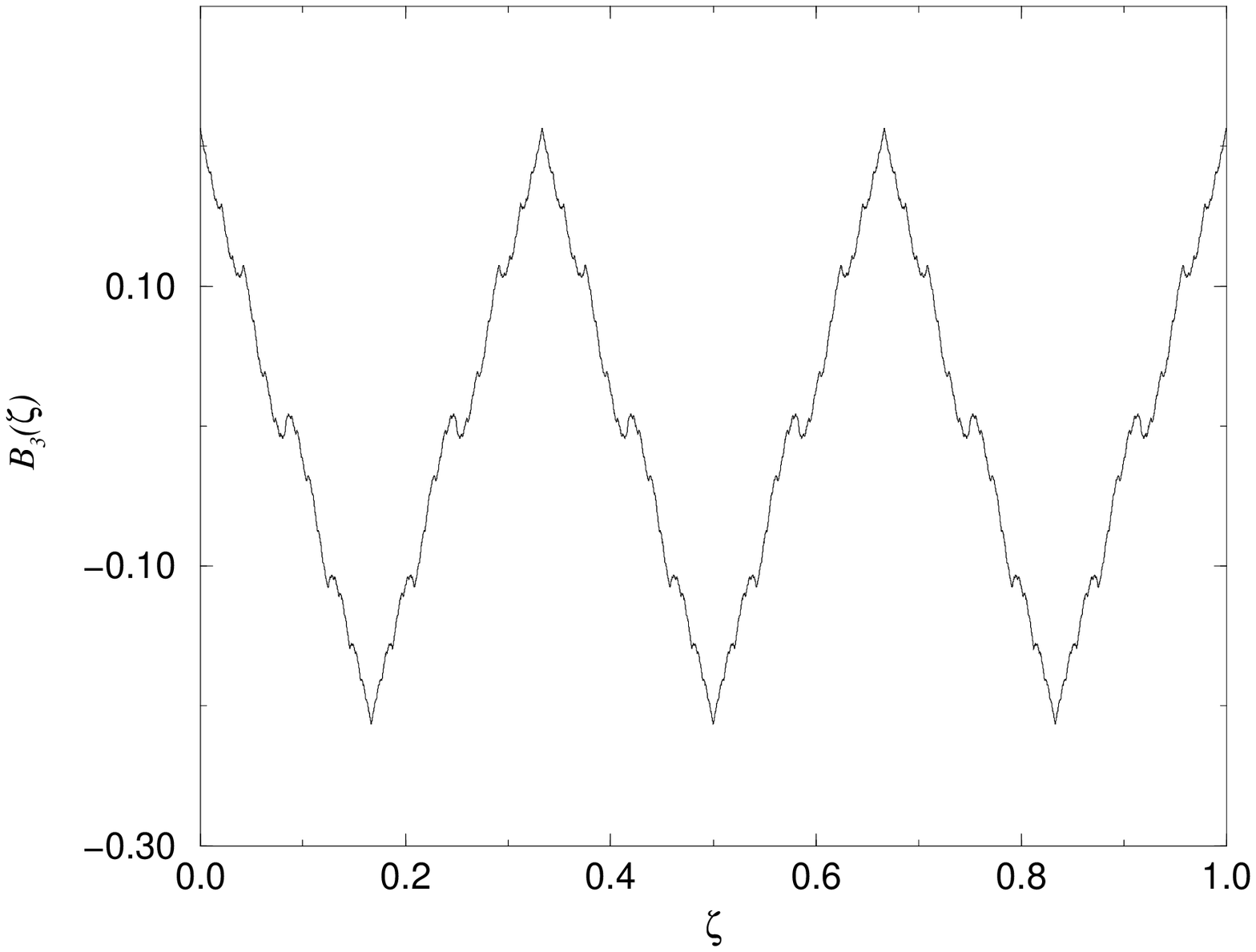,width=7cm}} 
\caption{\label{figSM3}{From top to bottom and left to right~: 
$A_k(\zeta)$ for $k = 1$, 2 and 3, and  
$B_k(\zeta)$ for  $k = 1$, 2 and 3, numerically computed from  
Eqs. (\ref{SMAk}-\ref{SMBk}), where $f^{(1)}$ was taken from Eq. (\ref{ex2})}} 
\end{figure}

Hence the series in Eqs. (\ref{SMAk}-\ref{SMBk}) are uniformly convergent  
trigonometric series,  so that $A_k$ and $B_k$ are both continuous functions  
of $\zeta$. We notice a striking similarity between the expressions of the  
Fourier 
coefficients $A_k$ and $B_k$ and Weierstrass functions \cite{weierstrass1895},  
which are continuous, nowhere differentiable functions. As stated in 
Appendix {\ref{appendWF}} below, the fractal dimension of these functions depends on the scaling 
properties of the coefficients in the trigonometric series. In particular, 
the dimension of the graphs of $A_k$ and $B_k$  
defined by Eqs. (\ref{SMAk}-\ref{SMBk}) is unity, since the  
coefficients in the series decay like $1/2^n$. 
Figure (\ref{figSM3}) displays the first few Fourier modes $A_k$ and $B_k$ as  
functions of $\zeta$, for $k = 1, 2, 3.$ Note that modes of different $k$'s 
have similar shapes but decaying amplitudes as $k$ increases. 
 
Consider again Eq. (\ref{SMFourier}), which, with the help of   
Eqs. (\ref{SMAk}-\ref{SMBk}), can alternatively be written in the form 
\begin{equation} 
\rho^{(1)}(x) = -\sum_{k = 1}^\infty \sum_{n = 0}^\infty 
g_{2^n k}\sin\left\{2\pi k \left[\left(2^{n + 1} - 1\right)\zeta -  
x\right]\right\}. 
\label{SMrho1} 
\end{equation} 
As we pointed out already, the dependence of $\rho^{(1)}$ upon $x$ and 
$\zeta$ is more complicated than the dependence of $A_k,B_k$ on $\zeta$.  
The set of its discontinuities is dense with respect to both arguments, 
even though $\rho^{(1)}$ takes finite values everywhere. 
Nevertheless, the explicit formal functional form allows us to 
calculate some quantities of interest which involve $\rho^{(1)}$ only under 
the integration sign. Such quantities are related to the Lyapunov 
exponent of the map and the transport properties of the same map 
``lifted'' onto the real line. This is the subject of the next section.  
 
\section{\label{SMlyaptrans}Lyapunov exponents and transport properties}

The positive Lyapunov exponent can be computed 
up to second order in $a$ with the help of  
Eqs. (\ref{SMrhox},\ref{SMFourier})~: 
\begin{eqnarray} 
\lambda_+(a, \zeta) &=& \int_0^1 dx \rho(x)\log\left[\left|\frac{d 
    \varphi_{a,\zeta}(x)}{dx}\right|\right] \\ \nonumber   
&=& \log(2) - a^2\left(\fr{1}{4} + \fr{4}{\pi}\sum_{k = 1}^\infty 
\fr{k B_k(\zeta)}{4 k^2 - 1}\right)
+ O(a^4). 
\label{SMplyap} 
\end{eqnarray} 
 
With the expression of $B_k(\zeta)$ given by Eq. (\ref{SMBk}),  
the summation in the last line can easily be computed numerically.  
 
\begin{figure}[hbtp] 
\centerline{\psfig{figure=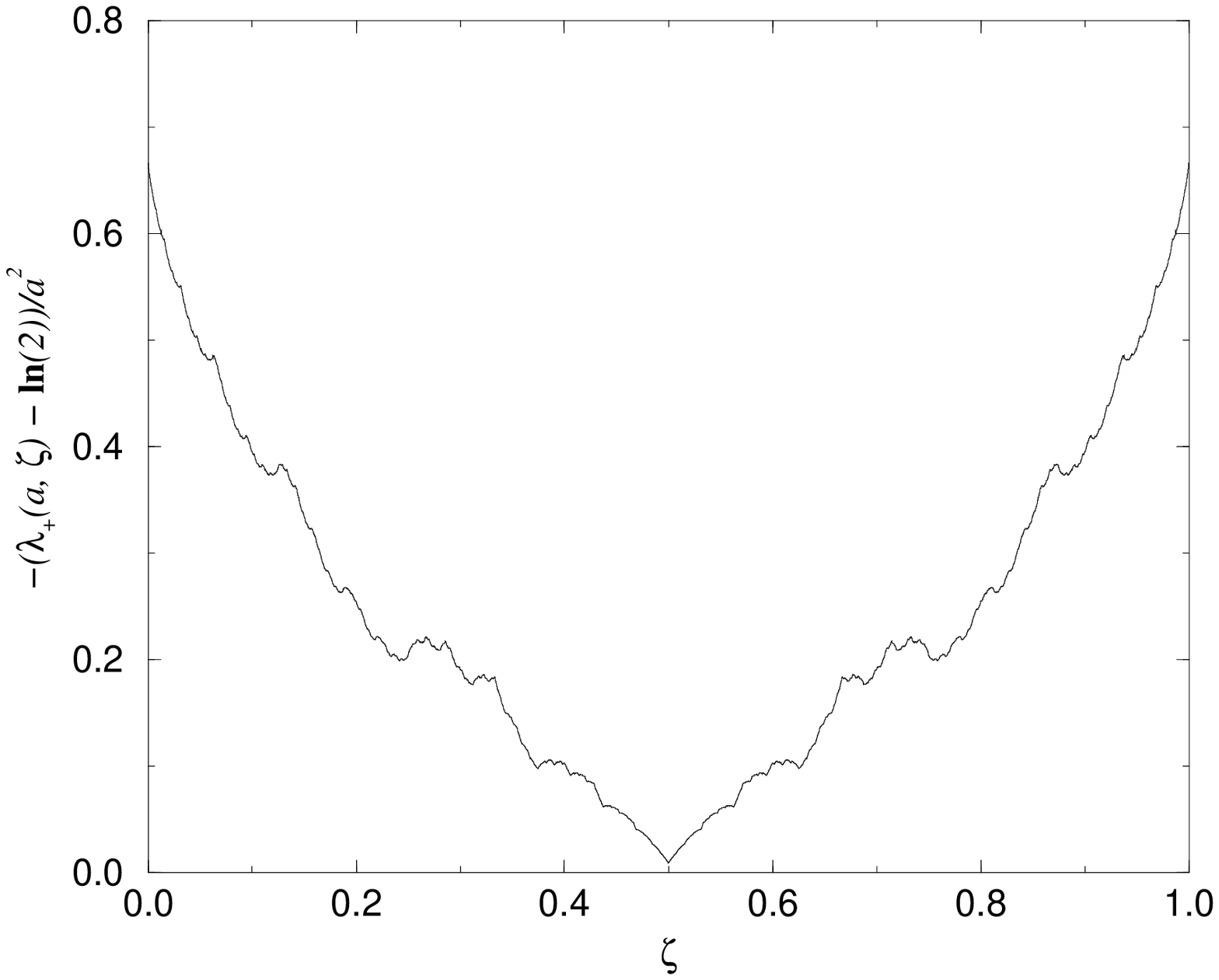,width=12cm}} 
\caption{\label{figSM4}{ Absolute value of the second order  
correction to  
$\lambda_{+}(a, \zeta)$, Eq. (\ref{SMplyap}). Here and in the next figure, 
$g$ is derived from Eq. (\ref{ex2}).}} 
\end{figure} 
Figure (\ref{figSM4}) shows the second order correction to  
$\lambda_+$ as a function of $\zeta.$  
The fact that it is a symmetric function 
with respect to $\zeta\rightarrow 1 - \zeta$ is a consequence 
of a symmetry of the map under $\zeta \rightarrow 1 - \zeta,\: a 
\rightarrow - a,\: x \rightarrow 1 - x$ and the fact that the  
positive Lyapunov exponent is a function only of the even powers of $a.$ 
 
\begin{figure}[htbp] 
\centerline{\psfig{figure=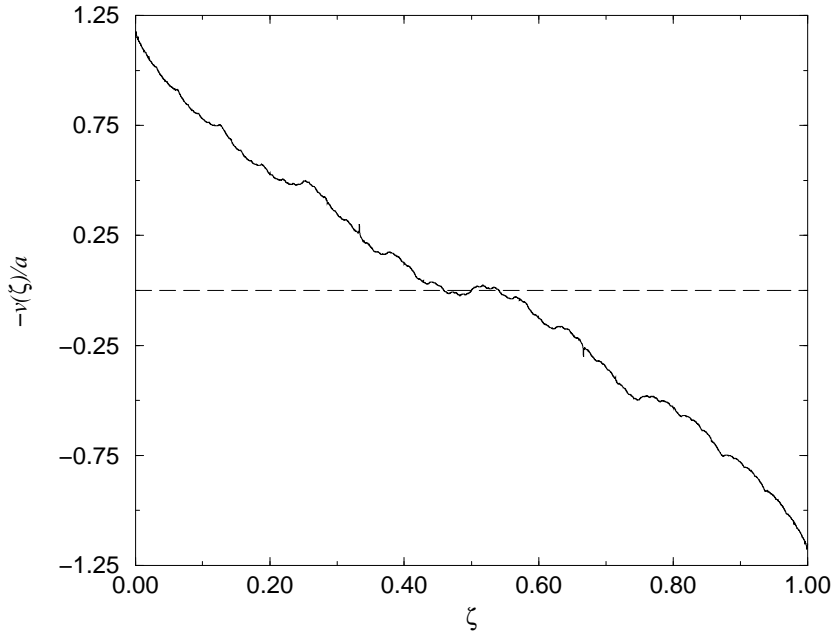,width=12cm}} 
\caption{\label{figSM5}{ Drift velocity to first order in  
$a$, Eq. (\ref{bavdrift}).}} 
\end{figure}

The stationary drift velocity measures the exchange of particles between the unit 
cells of the periodically extended version of $\varphi_{a,\zeta}$. In 
this extension, based on the multi-baker map, points are sent to 
corresponding points in the unit 
intervals to their right and left after each iteration of the 
map. Points in the interval $(f_{a}((1-\zeta)/2),f_{a}((2-\zeta)/2))$, mod$(1)$, 
are sent one unit interval to the right, and assigned a velocity $ 
v=+1$, while the remaining points in the interval $(0,1)$ 
are sent one unit to the left and assigned a velocity $v=-1$, with 
identical behavior in each unit interval of the real line. This is, in 
fact, the projection onto the $x$-axis of the nonlinear multi-baker 
map described in Appendix B (see Eq. (\ref{SMrotmb})). The 
expression for the stationary drift is now straightforward and yields 
\begin{eqnarray} 
v (a, \zeta) &=& -\int_0^{ f_{-a}((1 - \zeta)/2)} dx \rho(x)\nonumber\\ 
&&+ \int_{ f_{-a}((1 - \zeta)/2)}^{ f_{-a}((2 - \zeta)/2)}  
dx \rho(x) - \int_{ f_{-a}((2 - \zeta)/2)}^1 dx \rho(x).\nonumber\\ 
&=& -\fr{2a}{\pi}\Big\{\cos\left(\fr{\pi\zeta}{2}\right) - 
\sin\left(\fr{\pi\zeta}{2}\right)\nonumber\\ 
&& + \sum_{k\:{\rm odd}}\sum_{n = 0}^\infty  
\fr{g_{2^n k}}{k}\cos\left[\left(2^{n + 2} - 1\right)\pi k\zeta\right] 
\Big\} + O(a^3),\label{bavdrift} 
\end{eqnarray} 
Its graph is displayed in Fig. (\ref{figSM5}).  
The drift velocity is an odd function of $\zeta - 1/2$.  
An interesting feature of Fig. (\ref{figSM5}) is that the drift velocity  
is negative in a small interval about $\zeta = 1/2.$ This is another example  
of the negative current phenomenon discovered by Groeneveld in one  
dimensional maps, and discussed by Groeneveld, Klages, and coworkers  
\cite{groeneveld00,klages98,gaspard98b}, and it loosely 
corresponds to a negative electrical conductivity, where the current is in a  
direction opposite to that of the electric field, taken to be in the 
direction of $\zeta -1/2$. 
 
In Appendix  \ref{SMtwoparam} we present a generalization of our 
one-dimensional maps to two-dimensional maps, known as 
multi-baker  
maps. These maps have both a positive and a negative Lyapunov 
exponent. We believe that their sum  
obeys an identity relating the phase space contraction rate to the drift 
velocity, and is given as Eq. (\ref{driftlyap}). We do not yet have a rigorous 
proof of this identity, however.  
 
\section{\label{SMnonhyper}Non hyperbolic regime} 
 
Next, we describe some of the interesting features of the 
map $\varphi_{a,\zeta}$ for higher values of the field parameter. In 
Appendix \ref{appendTRP}, we present a classification of the non-linear 
maps $f_a$ satisfying the conditions listed in Sec. \ref{SMsrb}. We will 
restrict our attention to $f_a$ given by 
\begin{equation}\label{SMcurtaina} 
f_a(x) = \fr{2}{\pi}\arctan\left[\tan\left(\fr{\pi x}{2}\right) 
e^{- a}\right], 
\end{equation} 
which is derived in Appendix B. As was  
mentioned in \cite{gilbert99}, for $\zeta = 0,$ 
the map loses its hyperbolicity for $a > a_c = \log(2)$ and the  
origin becomes an attractive fixed point of the reduced map, so that, on a  
periodic lattice, the stationary state motion of particles is 
ballistic and moves toward decreasing $n$'s. 
 
Things change quite dramatically as $\zeta$ is increased. The origin 
is now mapped to $\zeta$ and, for $\zeta$ large enough, the orbit of the  
origin can  eventually become hyperbolic, even though the slope of  
$\varphi_{a,\zeta}$ 
at $x=0$ would be less than one. The fixed point is not the origin $x = 0$  
anymore  
and it becomes attractive for larger and larger values of $a$ as one 
increases $\zeta,$ i.~e. $a_c(\zeta),$ the value of the field parameter 
at which the fixed point becomes attractive, is a monotonically increasing 
function of $\zeta.$  
 
\begin{figure}[htbp] 
\centerline{\psfig{figure=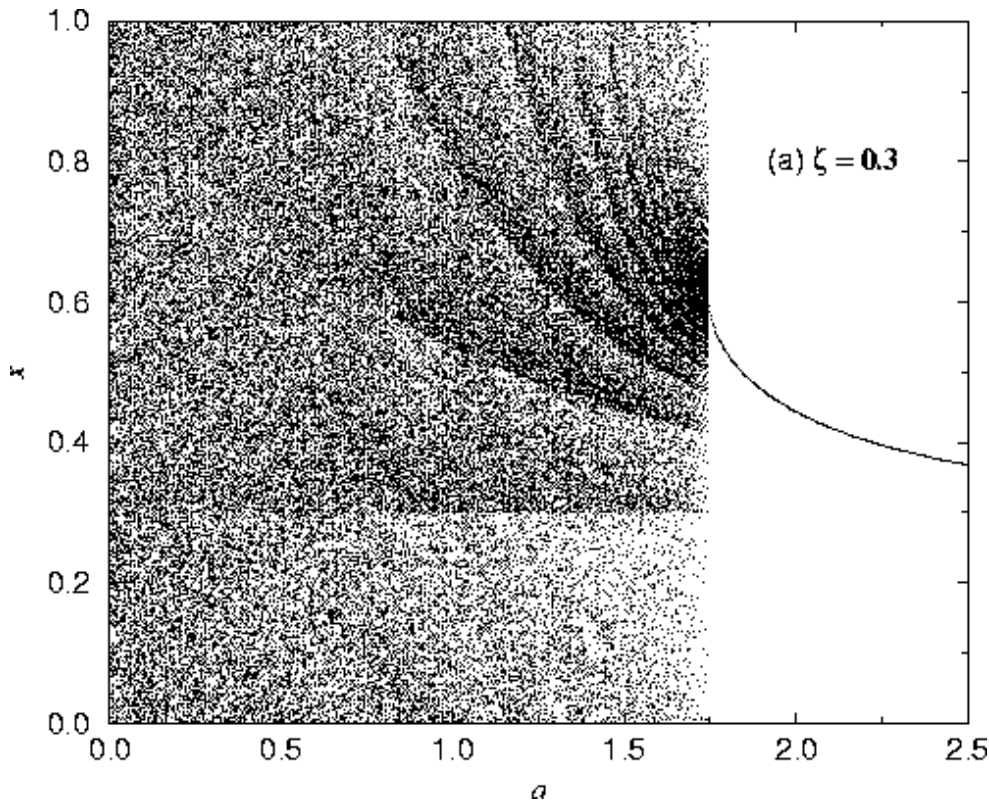,width=13cm}} 
\caption{\label{figSM6}{ The bifurcation diagrams of  
$\varphi_{a,  
\zeta}(x)$ for $\zeta = 0.3$ as a function of $a$.}} 
\end{figure}

Moreover, for fixed $\zeta$, stable periodic orbits can be observed  
in the range of $a$ between $\log(2)$ and $a_c,$ as shown in Fig. 
(\ref{figSM6}). A property of these windows is that they  
all have $0$ as a fixed point for the value of $a$ right before they  
bifurcate to chaotic windows. Based on this 
observation, it is possible to work out a condition for the presence 
of a window of stable period $n$ trajectories, namely that 
\begin{eqnarray} 
\varphi_{a,\zeta}^n(0) &=& 0,\nonumber\\ 
\left.\fr{d}{dx}\varphi_{a,\zeta}^n(x)\right|_{x = 0} &\leq& 1, 
\label{SMperiodn} 
\end{eqnarray} 
where the superscript $n$ denotes the $n_{\rm th}$ iterate. 
The system Eq. (\ref{SMperiodn}) can be studied in order to find the minimal value  
of $\zeta$ for which there exists such a stable period $n$ orbit and the  
corresponding value of $a$ at which it occurs.  
 
We illustrate this procedure for the simplest case of $n = 2.$  
From the top line of Eq. (\ref{SMperiodn}), we find 
\begin{equation}\label{SMafuncb} 
e^a = \fr{2\tan(\pi\zeta/4)}{[1 - \tan(\pi\zeta/4)]^2}, 
\end{equation} 
while the bottom line gives 
\begin{equation}\label{SMepabound} 
e^a \geq \fr{7 + \cos(\pi\zeta)}{1 + \cos(\pi\zeta)}. 
\end{equation} 
The minimal value of $\zeta$ satisfying Eqs.  
(\ref{SMafuncb}, \ref{SMepabound}) is $\zeta \simeq 0.522873,$ for which  
$a \simeq 1.005053.$ The existence of this periodic orbit is confirmed  
numerically. 
 
Higher $n$'s are found to start appearing at lower values of  
$\zeta,$ which  
means that, for a fixed $\zeta,$ all the periods $n$ greater than the first  
existing one will appear and they do so as $a$ increases until it 
reaches its critical value $a_c.$  
Such a succession of stable periodic windows with increasing periods is  
referred to as a period adding-bifurcation,  
\cite{levi90}.
Here stable periodic windows alternate with chaotic ones. The similarity  
of the bifurcations observed here with period-adding bifurcations of  
reversible circle maps can be explained qualitatively if one notices that the 
two branches of $\varphi_{a,\zeta}$ have very different widths. Indeed the 
first critical point $x=f_a(1/2)$ gets closer to $x=1$ as $a$ increases, so  
that the first branch is close to a reversible circle map, whereas the  
second branch has a very steep slope. In this sense, for a fixed large $a$, 
as we vary $\zeta$, we expect to see periodic orbits that remain on the first  
branch and thus correspond to stable windows, while the second branch accounts 
for the chaotic windows separating stable periodic windows. 
 
\begin{figure}[htbp] 
\centerline{\psfig{figure=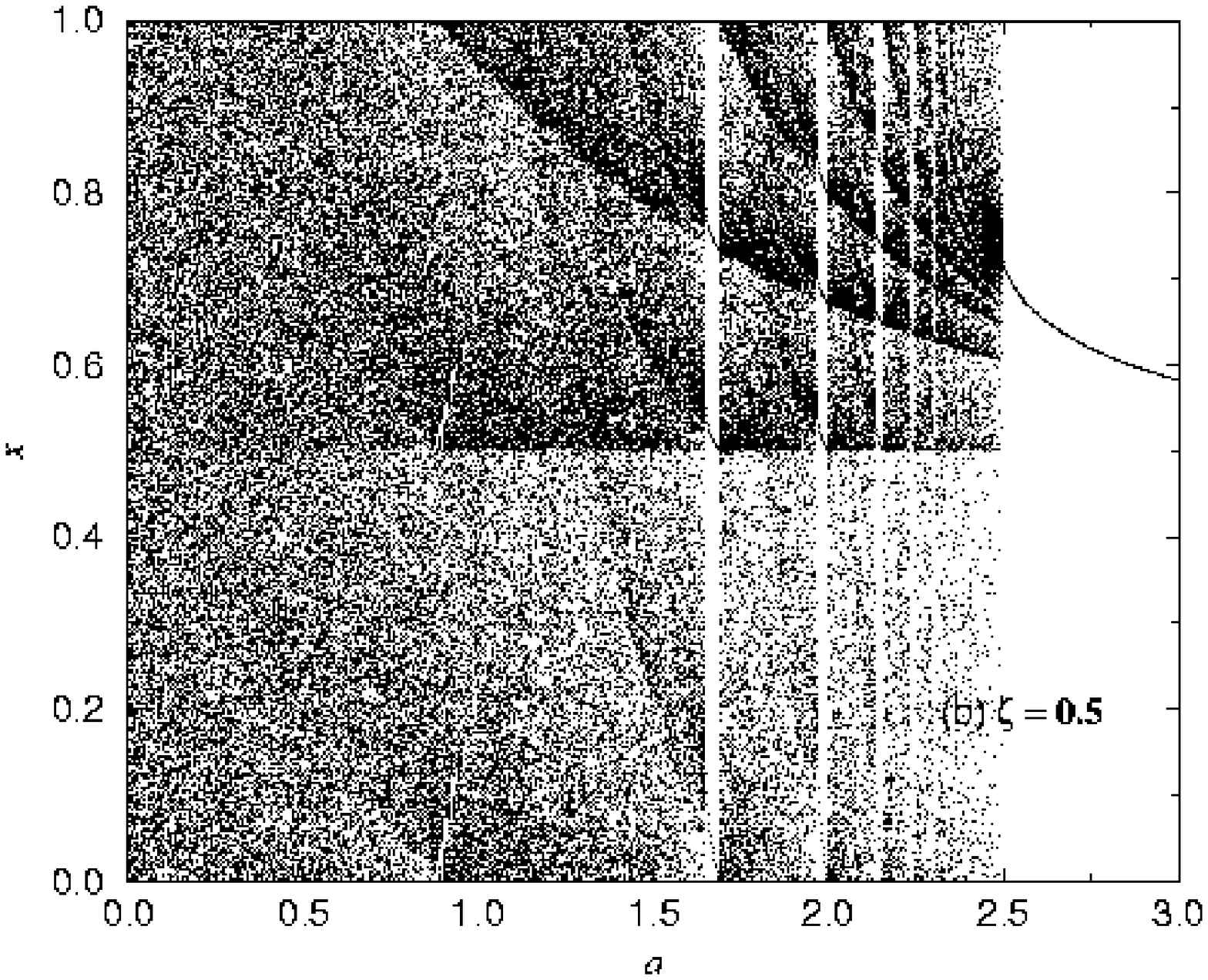,width=13cm}} 
\caption{\label{figSM7}{ Same as in Fig. (\ref{figSM6})  
for $\zeta = 0.5.$}} 
\end{figure} 
 
\begin{figure}[htbp] 
\centerline{\psfig{figure=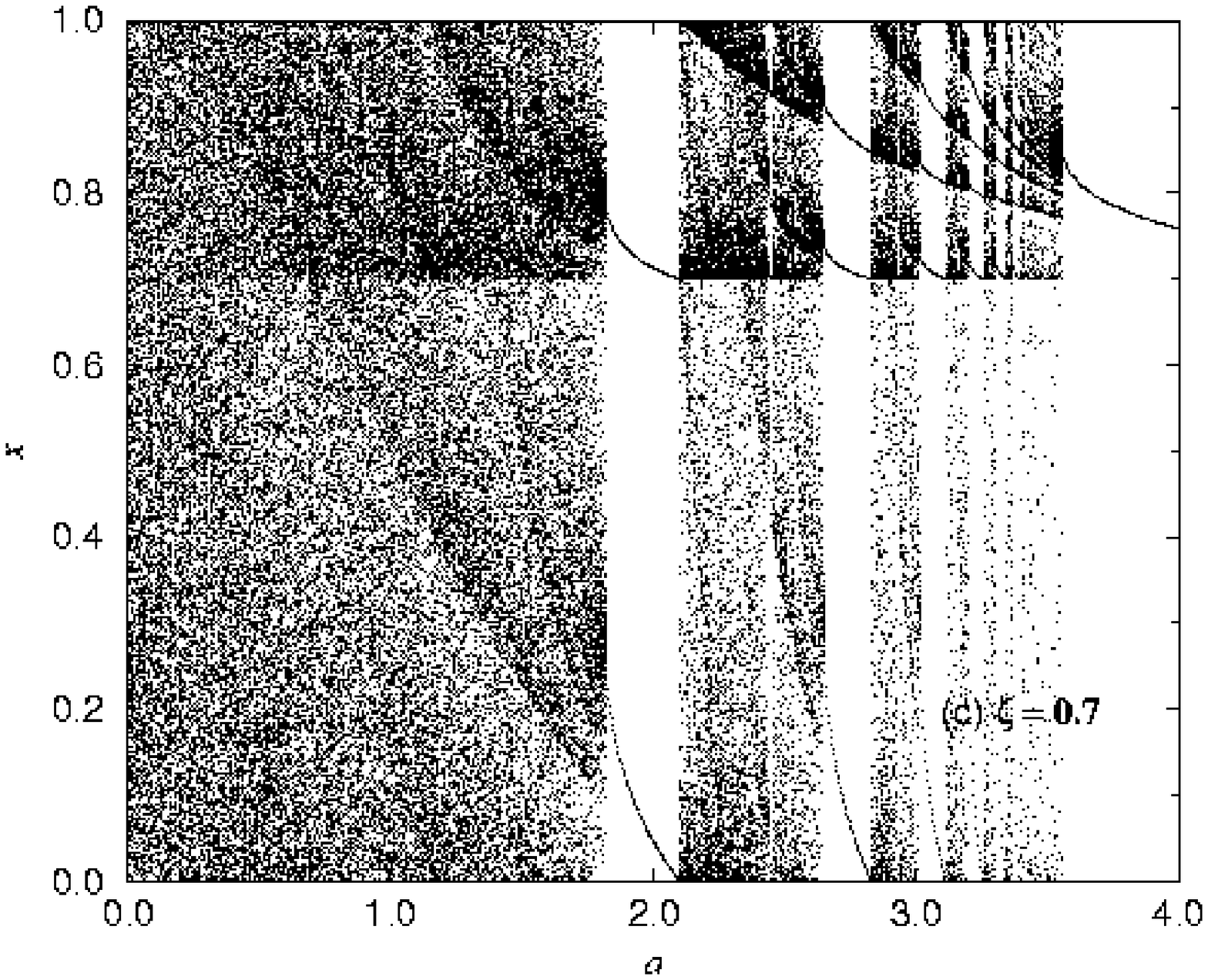,width=13cm}} 
\caption{\label{figSM8}{ Same as in Fig. (\ref{figSM6})  
for $\zeta = 0.7.$}} 
\end{figure} 
 
We illustrate this in Figs. (\ref{figSM6}-\ref{figSM8}) for the case of  
$\zeta = 0.3,$ 0.5, and 0.7.  
In the first example,  
$\zeta = 0.3,$ stable periodic windows are 
barely visible but one can check the existence of a period 6 window at 
$a \simeq 1.4449$ whose width is $O(10^{-4}).$ As $\zeta$ is increased 
to 0.5, windows get larger and, at $a = 1.6492$ a period $3$ window  
starts that ends at $a = 1.6881,$ then a period $4$ window arises at  
$a = 1.9678$ that ends at $a = 1.9985,$ and so on. But this is not  
the end of the story~: besides the main sequence, there are a number of  
subsequences, such as $2n + 1,$ $3n + 1,$ $3n + 2,$ etc., some of which are 
clearly distinguishable on the third example, $\zeta = 0.7.$ The higher  
$\zeta,$ the greater the number of such subsequences. 
\section{Conclusions and Discussion} 
 
In this paper we considered a class of two-parameter singular circle maps. 
One of these two parameters, denoted $a$, measures the intensity of the 
non-linearity. In the generalization of the non-linear baker map considered 
in Appendix \ref{SMtwoparam}, it plays the role of a thermostatted electric field. 
The other parameter, denoted $\zeta$, represents an external field, similar 
in some ways to a magnetic field.  
We found that the map exhibits both hyperbolic and  
non-hyperbolic behaviors, and that for small non-linearities, we could 
determine analytically the SRB measure of the fractal attractor to   
first order in $a$ and the first non zero correction, of order  
$a^{2},$ to the positive Lyapunov exponent. For large enough $a$ we  
observed a transition to non-hyperbolic behavior with a period-adding  
bifurcation sequence leading to a stable fixed point for large enough  
$a$. 
 
Thus, the rotation parameter leads to a much more complicated dynamical  
behavior of the map, including a region of ``negative currents'' similar to  
those found before by Groeneveld, Klages, and coworkers  
\cite{groeneveld00,klages98,gaspard98b}. 
 
The results of this paper lead us to the following observation~: 
The dynamical systems studied here are, in their hyperbolic regions,  
 not Anosov diffeomorphisms. We find that the SRB measure  
of the fractal attractor is not a differentiable function of the parameter  
$\zeta$, as it would be if the map were an Anosov diffeomorphism (i.~e. did 
not have a singularity) 
\cite{ruelle97b,gallavotti97}.  
Instead, it is a singular function of that parameter.  
As hyperbolic systems which are not Anosov occur frequently in models of  
physical systems \cite{gallavotti95}
it is not unreasonable to suppose that  
there will exist situations where non-differentiable SRB measures occur in  
physical systems. Our work suggests, at least, that this might very well be  
the case in Lorentz gases or other discontinuous systems in the presence of  
external fields, where singularities of the flow would be rotated as the 
field is varied. 
 
\section*{Acknowledgments} 
The authors wish to thank Brian Hunt, Celso Grebogi, Edward Ott, 
Helena Nusse, Rainer Klages, Lamberto Rondoni, Karol \.Zyczkowski, 
Christian Maes, Gary Morriss, Nikola\"{\i} Chernov, Daniel Wojcik,  
Mihir Arjunwadkar, Gerald Edgar, Dan Mauldin and Loren Pitt for helpful  
discussions.  
J. R. D. wishes to acknowledge support from the National Science Foundation 
under grant PHY-98-20428. TG acknowledges financial support from the 
European Union under contract number HPRN-CT-2000-00162.

\begin{appendix} 
\section{Weierstrass function\label{appendWF}} 
For $\lambda>1$ and $1<s<2$, the Weierstrass function \cite{weierstrass1895}  
can be defined as  
\begin{equation} 
W_{\lambda,s}\::\:[0,1]\longrightarrow\mathbf{R},\quad t\longrightarrow  
\sum_{k=1}^\infty 
\lambda^{(s-2)k}\sin(\lambda^k t)\ . 
\end{equation} 
It is a nowhere differentiable continuous function \cite{hardy16}. 
 
As stated by Falconer \cite{falconer90}, provided $\lambda$ is large enough, 
the box counting dimension of the graph of this function is 
\begin{equation} 
\mbox{dim}_B {\mathcal G} (W_{\lambda,s}) =s \ . 
\end{equation} 
In general, the function 
\begin{equation} 
F_{\lambda,s}(t)=\sum_{k=1}^\infty\lambda^{(s-2)k}g(\lambda^k t) 
\end{equation} 
has box counting dimension $s$. 
This constitutes an upper bound for the Hausdorff dimension. However, it is 
commonly believed \cite{edgar93} that the Hausdorff dimension is exactly 
equal to $s$. 
 
The case $s=1$, which is relevant to the Fourier modes given by Eqs.  
(\ref{SMAk}-\ref{SMBk}), is borderline. In this case, despite the nowhere  
differentiability  
of the Weierstrass function, it is understood that its Hausdorff dimension is equal  
to 1  
\cite{edgar01,mauldin01}. 
 
\section[Non-linear baker maps] 
{\label{SMtwoparam}Non-linear baker maps} 
 
Motivated by the analytical study of non-equilibrium stationary states of  
thermos\-ta\-ted field driven systems, Gilbert {\em et al.} introduced in  
\cite{gilbert99} the one parameter family of maps 
\begin{equation} 
\label{SMfieldbaker} 
M_a(n, x, y) = 
({\bf 1}, {\bf 1}, f_a)\circ M_0 \circ 
({\bf 1}, f_a, {\bf 1})\:(n, x, y), 
\end{equation} 
where ${\bf 1}$ denotes the identity operator, acting either on integers or on 
points of the unit circle, $M_0$ is the usual multi-baker map  
\cite{gaspard92,tasaki95,gaspard98}, 
\begin{equation} 
M_0(n, x, y) =\left\{ 
\begin{array}{l@{\quad}l} 
(n - 1, 2x, \fr{y}{2}),& 0 \leq x < \frac{1}{2},\\ 
(n + 1, 2x - 1, \fr{y + 1}{2}),&\quad \frac{1}{2} \leq x < 1, 
\end{array} 
\right. 
\label{SMmb} 
\end{equation} 
with variables $n \in {\bf Z}$ and $ (x,y) \in [0, 1]^2$. Here 
$f_a$ is a non-linear perturbation of the unit interval onto itself, which 
models the action of a thermostated external field, 
\begin{equation}\label{SMcurtain} 
f_a(x) = \fr{2}{\pi}\arctan\left[\tan\left(\fr{\pi x}{2}\right) 
e^{- a}\right]. 
\end{equation} 
 
This model is a caricature of the two-dimensional periodic thermostated  
Lorentz gas with an applied electric field 
\cite{lloyd95}
Here the variable $\pi x$ plays the role of the angle that the velocity  
makes with the direction of the electric field. 
 
We generalize Eq. (\ref{SMfieldbaker}) by introducing, in the multi-baker 
map, Eq. (\ref{SMmb}), a new parameter, $\zeta,$ defined modulo 1, acting 
on the unit square, which 
shifts points by an angle $\zeta$ in the $x$-direction and acts on the 
$y$-direction so as to guarantee a weakened form of reversibility, see  
Eq. (\ref{SMtrev2pb}) below. 
 
The new multi-baker map, Fig. (\ref{figSM1}), 
constructed so as to reduce 
to the multi-baker map, Eq. (\ref{SMmb}), for $\zeta=0$ and $1$, reads~: 
\begin{equation}\label{SMrotmb} 
M_{0, \zeta}(n, x, y) = \left\{ 
\begin{array}{l@{\quad}l} 
(n - 1, 2 x + \zeta, \fr{y + 2 - \zeta}{2}),& 
0 \leq x < \fr{1 - \zeta}{2},\\ 
&\quad 0\leq y < \zeta,\\ 
(n - 1, 2 x + \zeta, \fr{y - \zeta}{2}), 
&0 \leq x < \fr{1 - \zeta}{2},\\ 
&\quad \zeta\leq y < 1,\\ 
(n + 1, 2 x +\zeta - 1, \fr{y + 1 - \zeta}{2}),& 
\hspace{-.25cm}\fr{1 - \zeta}{2} \leq x < \fr{2 - \zeta}{2},\\ 
&\quad \:0\leq y <1,\\ 
(n - 1, 2 x +\zeta - 2, \fr{y + 2 - \zeta}{2}),& 
\fr{2 - \zeta}{2} \leq x < 1,\\ 
&\quad 0\leq y < \zeta,\\ 
(n - 1, 2 x +\zeta - 2, \fr{y - \zeta}{2}), 
&\fr{2 - \zeta}{2} \leq x < 1,\\ 
&\quad \zeta\leq y < 1.\\ 
\end{array}\right. 
\end{equation} 
 
\begin{figure}[htbp] 
\centerline{\psfig{figure=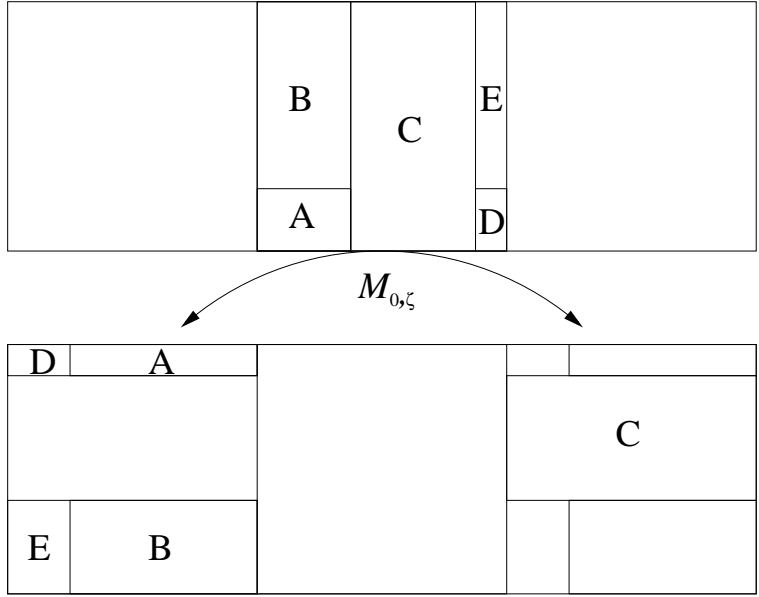,width=10cm}} 
\caption{\label{figSM1}{ $M_{0,\zeta}$ for $\zeta = 0.25.$}} 
\end{figure}

The action of the time reversal operator $T,$ $T(x, y) = (1 - y, 1 - x),$ on 
this map can be seen to give 
\begin{equation}\label{SMtrev2pb} 
T \circ M_{0, \zeta} \circ T = M_{0, 1 - \zeta}^{- 1}. 
\end{equation} 
 
Thus, under time reversal, the map (\ref{SMrotmb}) becomes the inverse 
of 
$M_{0,1-\zeta}$ instead of the inverse of $M_{0,\zeta}$. This property is 
reminiscent of the effect of time reversal on an external magnetic field,  
which changes the sign of the field.  
 
We now define the two parameter map $M_{a,\zeta}(n,x,y),$ the same way 
as $M_a$ in Eq. (\ref{SMfieldbaker}), by 
\begin{equation} 
\label{SM2parambaker} 
M_{a,\zeta}(n, x, y) = 
({\bf 1}, {\bf 1},  f_a)\circ M_{0, \zeta} \circ 
({\bf 1},  f_a, {\bf 1})\:(n, x, y), 
\end{equation} 
 
Despite the lack of a precise reversibility, it is interesting to ask 
whether the the usual result relating the irreversible entropy production 
to the sum of the Lyapunov exponents still holds~: 
\begin{equation}\label{driftlyap} 
\fr{{\overline v}^2}{D} = - (\lambda_+ + \lambda_-). 
\end{equation} 
We believe it is the case here. This view is motivated by the symmetries 
present. However a definite answer based upon rigorous arguments has yet to 
be reached.  
\section{Classification of time-reversible perturbations  
\label{appendTRP}} 
In this appendix we present a classification of the perturbations $f_a(x)$,  
with fixed points at $x=0$ and $x=1$, that  
satisfy the time-reversibility properties, Eqs. (\ref{addfa}-\ref{trfa}).  
In order to find explicit forms of perturbations, it is natural to assume 
factorizability of the $x$ and $a$ dependences.  
Thus consider the two classes 
\begin{eqnarray} 
f_a(x)&=& g^{-1}\circ[g(x) q(a)]\ ,\label{case1}\\ 
f_a(x)&=& h^{-1}\circ[h(x) + r(a)]\ .\label{case2} 
\end{eqnarray} 
Here the parameter functions $q$ and $r$ are any functions of $a$ with the  
respective properties, 
\begin{eqnarray} 
q(0) &=& 1\ , \label{propqa}\\ 
q(a)^n&=&q(n a), \label{propqb}\\ 
q(a) q(-a) &=& 1\ ,\label{propqc}\\ 
r(0) &=& 0\ ,\label{propra}\\ 
n r(a) &=& r(n a)\ , \label{proprb}\\ 
r(a) + r(-a) &=& 0\ ,\label{proprc} 
\end{eqnarray} 
where $n$ is a positive integer. In particular these conditions imply the  
property Eq. (\ref{invfa}) for $f_a$. We note 
that any function $r$ of the second class defines a function $q$ of the first 
class by exponentiation, $q(a)=\exp[r(a)]$. The only function $r$ satisfying 
Eqs. (\ref{propra}-\ref{proprc}) is a linear function $r(a) = c a$, where  
$c$ is an arbitrary constant that can be taken to be $c=- 1$ without loss 
of generality, since $a$ is allowed to take any real value. Likewise  
$q(a)=\exp(-a)$ is the only function that needs to be considered. 
 
The reversal symmetry, Eq. (\ref{trfa}), is satisfied provided 
\begin{eqnarray} 
1 - g^{-1}\circ[g(x) q(a)]&=&g^{-1}\circ[g(1-x)/q(a)]\ , \label{propg}\\ 
1 - h^{-1}\circ[h(x) + r(a)]&=&h^{-1}\circ[h(1-x)-r(a)]\ . \label{proph} 
\end{eqnarray} 
A class of functions verifying Eq. (\ref{propg}) is the set of functions 
$g$ with the properties 
\begin{eqnarray} 
g(x)&=&\frac{1}{g(1-x)}, \label{propga}\\ 
g^{-1}(x)&=&1 - g^{-1}\big(\frac{1}{x}\big)\ ,\label{propgb} 
\end{eqnarray} 
Notice that the requirement that $x=0$,$1$ be fixed points of $f_a$ implies 
that $g$ is either zero or plus or minus infinity at these points. Such 
examples are $g(x) = \tan(\pi x/2)$, $\cot(\pi x/2)$, which define identical 
$f_a$'s up to a change of sign of $a$.  
 
Similarly a class of functions verifying Eq. (\ref{proph}) is defined by 
\begin{eqnarray} 
h(x) &=& -h(1-x)\ ,\label{propha}\\ 
h^{-1}(x)&=&1 - h^{-1}(-x)\ , \label{prophb} 
\end{eqnarray} 
examples of which are $h(x) = \cot(\pi x)$, $\cos(\pi x)$. Note however that 
$\cos(\pi x)$ is not consistent with the requirement that $x=0$, $1$ remain 
fixed points of $f_a$. Moreover the choice $\cot(\pi x)$ yields a perturbation 
$f_a$ with smooth derivative at the origin, $f_a'(0)=f_a'(1)=1$. This can be 
easily seen to yield a differentiable density $\rho^{(1)}$. 
 
We therefore argue that the choice of $f_a$ given by Eq. (\ref{SMcurtaina}) is 
generic. Other forms of perturbations such as polynomial in powers of $a$ can  
be constructed, but they do not have a complete form and can only be  
considered  
for values of $a$ small enough with respect to the degree of the polynomial  
approximation. Thus the non-hyperbolic regime, where $a$ becomes large, can 
only be studied with the example given in Sec. \ref{SMnonhyper}. 
 
 
\end{appendix}

\end{document}